# Radon Concentration Potential in Bibala Municipality Water: Consequences for Public Consumption

Joaquim Kessongo[a,c], Yoenls Bahu[a,c], Margarida Inácio[a,c], Pedro Almeida[d,e], Luis Peralta[c,e], Sandra Soares[a,b,c]

[a]*Departamento de Física, Faculdade de Ciências da Universidade da Beira Interior, Covilhã, Portugal*
[b]*Centro de Matemática e aplicações da Universidade da Beira Interior, Covilhã, Portugal*
[c]*Laboratório de instrumentação e Física Experimental de Partículas, Lisboa, Portugal*
[d]*Departamento de Engenharia Civil e Arquitetura, Faculdade de Engenharia da Universidade da Beira Interior, Covilhã, Portugal*
[e]*Faculdade de Ciências da Universidade de Lisboa, Lisboa, Portugal*

Abstract

The primary motivation for this work is the evaluation of the radon concentration in portable water for human consumption in Bibala, a municipality in Angola, where granitic rocks are common, and contain a high concentration of uranium that can be mobilized in underground water. Radon is the largest contributor of radioactive pollution in underground water. Its concentration in water, represents a public health risk due to the fact that the gas can easily escape into the air, adding to the total indoor concentration of radon. On the other hand, ingestion of water with a high radon concentration represents an additional risk to the stomach. Measurements of radon concentration, in Bibala municipality's water, were performed on 16 samples obtained from wells of various depths and analyzed with DURRIDGEs' RAD7 equipment. Measured concentrations are in the range from 39.5 to 202 Bq L$^{-1}$, with 2 of the recovered samples presenting values over 100 Bq L$^{-1}$.

1. Introduction

Since the beginning of the world human beings have been exposed to natural radiation deriving from the earth as well as from outside. Naturally occurring radioactive materials (NORM) are present in the earth crust, in building materials, in various types of food and drinks and can be dissolved in very low concentrations during normal interactions between water and rock or soil [1].

Radon $^{222}$Rn, is a noble and naturally occurring radioactive gas that is present in the environment, including air, soil and water. This is the only gaseous element produced during the uranium, $^{238}$U, decay chain. Radon gas and its progeny's concentration survey has become of extreme relevance since it was recognized in 1988 by the World Health Organization (WHO) as carcinogenic, the second leading cause of lung cancer after tobacco smoke [2].

Radon being gaseous and chemically inert doesn't react with other elements and because of its high mobility can escape from the soil easily. Although it has harmful effects in closed spaces, its presence in the environment is not perceived by humans and is normally used as an atmospheric tracer because of its relatively long half-life of 3.82 day [3].

Approximately 85% of the radiation dose received by the population has its origins in natural radioactive sources, where about half of it is received by the inhalation or ingestion of radon [4]. Radon isotopes can escape into the atmosphere by the diffusion process and its atmospheric abundance is related to its exhalation rate from soils which increase wind velocity [5]. When radon is inhaled the alpha particles emitted by its short-lived decay products ($^{218}$Po and $^{214}$Po) deposit all of its energy into the cells of the respiratory system's air-ways, causing damage. Fractures in rocks and fissures in soils allow rapid propagation of radon. In water radon mobility is lower than in the air and the distance traveled until it decays is usually no more than 2.5 cm. Consequently, in dry regions with permeable soils and rocks, for instance on mountains slopes, radon levels tend to be higher than nearby flat and wet soils [6].

Radon exhaled from rocks and soils into the atmosphere infiltrates dwellings and other buildings through three main processes: air pressure difference between soil and the building; existence of fissures in the foundations; increased permeability around the foundations. The presence of radon in water comes from two different sources: decay of $^{226}$Ra solution in the water; direct exhalation of radon from $^{226}$Ra decay present on the aquifer rock structure [6].

In regions where radon could be present in groundwater, if drinking water comes from wells, it may contribute as an additional source of airborne indoor radon because it may subsequently be inhaled. However when this water is ingested it is well documented that radon is not immediately transferred to the blood and then exits the body. The remaining radon in the stomach, due to this contaminated water, contributes to the stomach tissues being exposed to radon and this can imply an increased risk of stomach cancer [7].

The effects of radon ingestion are still not fully understood. Nevertheless, approximately one percent of radon exposure is due to domestic activities leading to the release of radon from water, like showering [8].

Most residences in Bibala where water samples were collected are served by underground pumping wells. Even though most of the radon is released into the atmosphere, it is necessary to evaluate its concentration. Already at the maximum recommended concentration by WHO of 100 Bq L$^{-1}$ radon gives a significant contribution to the increase of atmospheric indoor radon concentration.

In this paper, results of radon concentration in water samples collected between 2017 and 2018 in Bibala municipality (Angola) are presented. This is a pioneering study in the Bibala region, and the results are of significance to public health concerns.

2. Materials and methods

*2.1. Location of study area and geology setting*

The uranium and radium concentration in groundwater depends not only on the geological conditions but also on the lithology and geomorphology of the country. This water could move through and over the soil and rocks and can dissolve those natural radioactive heavy metals. When these elements are present in drinking water, they can build up in the body producing harmful effects for human health. At the same time, when radon is released from this drinking water it contributes to the increase of the indoor airborne radon concentration [9].

This study focuses on the extreme southeast of Angola in Bibala municipality located on the inland of Namibe province. It has a surface area of 7612 km$^2$ and according to Census from 2014 counted a total population of 55.399 inhabitants. It is limited on the north by Camucuio municipality, to the east by the municipalities of Quilengues, Cacula, Lubango and Humpata, to the south by Virei municipality and to the west by Moçamedes municipality [10]. The representative map of Bibala is shown in Figure 1 .

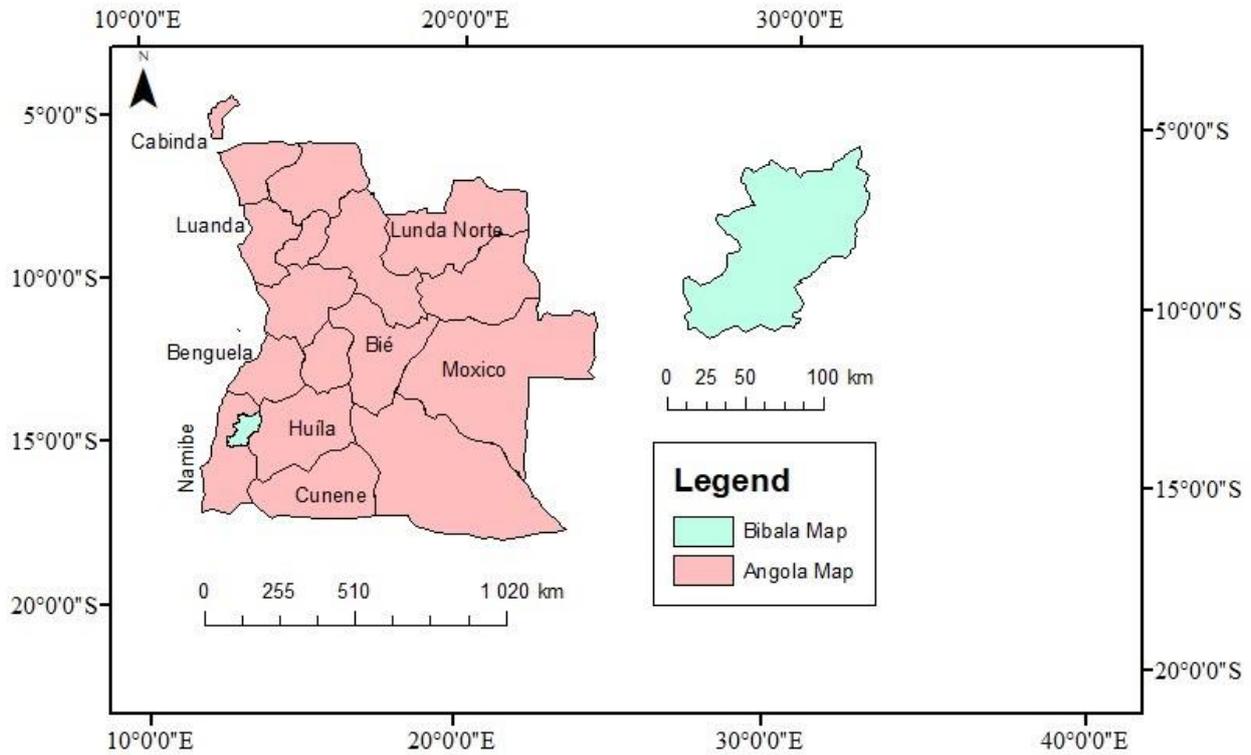

Figure 1. Bibala municipality, Angola, representative map (designed with ESRI's ARCGIS).

The climate in the Bibala region is semi-arid where the average annual temperature is around 23.7 °C influenced by the cold stream of Benguela and by morphology where inselbergs are abundant reaching 2500 m height. Relative humidity varies between 40 and 70% [10].

The Bibala's region geological complex presents granitic rocks generally of porphyroid texture and coarse matrix, calcium-alkali rich feldspars, quartz and microcline with an enrichment in FeMg minerals and variable uranium concentration. These occur between the areas of Moçamedes, Caraculo, Munhino and Luchipa [11].

### 2.2. Sampling

In this work, 26 water samples were carefully collected from 16 collection locations in the municipality of Bibala, during the Summer of 2017 and 2018. Radon gas is soluble in water and as its solubility decreases with an increase in temperature and can escape rapidly from the water into air. When the water is in plumbing, the radon concentration is lower than it is beneath the surface of the ground. For this reason, when we want to measure its concentration it is necessary to take all of the water that is held in the pipes. In order to preserve the quality of the sampling, the water was collected after approximately two minutes of pumping and sealed immediately to avoid radon escaping from them. To collect the water, 40 mL capacity glass vials were used. These special glass vials assure that the radon loss by degassing is as low as possible [12]. The collected samples were transported by plane to the Radon Exposure Effects Studies Laboratory (LabExpoRad) at Covilhã, Portugal for analysis.

### 2.3. Radon activity measurements

The laboratory measurements, of the radon-in-water concentration were performed with the RAD7 equipment using the $H_2O$ option, with a specific procedure that provide a direct reading of the water sample radon concentration. The measurement techniques are described in various documents [8]. In Figure 2 a schematic representation of the setup, as we can perceive, the radon activity concentration was measured in a closed circuit [13].

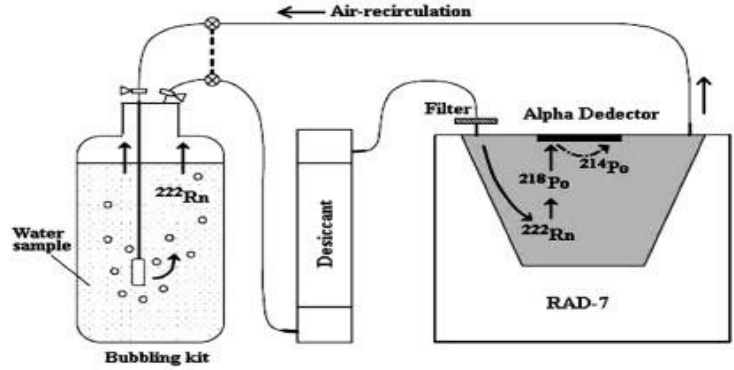

Figure 2. Schematic representation of radon in water concentration setup.

Radon expelled from the water sample by an aeration system enters through the detector device however, this inert gaseous alpha-emitter does not stick to or react with any materials. After decaying, radon releases 5.49 MeV alpha particles and transforms to polonium, $^{218}$Po, a short-lived isotope that emits an alpha particle of 6.00 MeV when it decays. Another important short-lived isotope is $^{214}$Po, that release an alpha particle of 7.69 MeV when it decays into lead, $^{210}$Pb. After decaying, the short-lived radon progeny, polonium nucleus, sticks to the active surface of the detector producing an electrical signal and, from the count rate of the detected nucleus. The radon monitor RAD7 uses a silicon solid-state detector to estimate the radon activity concentration, that is expressed in Bq m$^{-3}$ or Bq L$^{-1}$ with an uncertainty of 2σ [14].

*2.4. Estimation of annual committed effective dose by ingestion*

Radon is the principal natural radioactive agent that the world's population are exposed to on a daily basis. Its inhalation accounts, on average, for approximately fifty percent of the effective dose, in human beings. However, when radon is present in drinking water, it can escape into indoor air. Because of this fact, as the risk of lung cancer due to inhalation increases, the dose due to radon should be estimated in two ways: the dose from ingestion and the dose from inhalation.

To evaluate the radiological effects due to the ingestion of radon dissolved in drinking water, we need to calculate the annual committed effective dose, expressed in Sv y$^{-1}$, received by the population, using the equation [15] [16]:

$$D_{ing} = K_w \times C_{Rn} \times V \qquad (1)$$

where $K_w$ is the ingesting dose conversion factor for radon for ICRP age groups in (Sv Bq$^{-1}$) [17], $C_{Rn}$ is the radon activity concentration (Bq L$^{-1}$) and $V$ is the annual water consumption rate L y$^{-1}$ [18]. To estimate this dose, in this work, we used the values listed in Table 1.

Table 1. ICRP dose conversions factors and Annual water consumption

| Group ID | groups | $K_w$ (Sv Bq$^{-1}$) | ( L y$^{-1}$) |
|---|---|---|---|
| G I | Infants | 7 × 10$^{-8}$ | 150 |
| G II | Children | 2 × 10$^{-8}$ | 350 |
| G III | Adults | 1 × 10$^{-8}$ | 500 |

When radon enters the human body, its short-lived progeny will decay, not only inside the stomach but also in the lungs. The annual effective dose to the lungs from radon in water, were calculated, using the parameters established in the UNSCEAR report the dose conversion factor $K'_w = 9 \times 10^{-9}$ Sv Bq$^{-1}$ h$^{-1}$ m$^3$, the equilibrium factor between radon and its progeny F = 0.4, the radon transfer form water to air coefficient R = $10^{-4}$ and the average indoor occupancy time factor t = 7000 h y$^{-1}$, in the equation [18]:

$$D_{inh} = K'_w \times C_{Rn} \times F \times R \times t \qquad (2)$$

where $C_{Rn}$ is the radon concentration per cubic meter.

3. Results and Discussion

In order to better identify and map the region where drinking water presents more radon concentration the GPS coordinates were registered and presented in Table 2. Figure 3 represents the geographic distribution of radon in water activity concentration values based on the collected samples on a simple schematic representation.

Table 2. GPS coordinates of the collection locations of the water samples obtained during the summer of 2017 and 2018.

| S/N | Name of the water source | Sample ID | Latitude | Longitude |
|---|---|---|---|---|
| 1 | Kipamba | S1 | -14.721667 | 13.387778 |
| 2 | Mahita | S2 | -14.770833 | 13.346111 |
| 2 | Bomba | S3 | -14.747500 | 13.356667 |
| 4 | Matadouro | S4 | -14.747222 | 13.358333 |
| 5 | Matuco | S5 | -14.757222 | 13.353056 |
| 6 | Jamba | S6 | -14.473333 | 13.466111 |
| 7 | Camupupa | S7 | -14.525833 | 13.428050 |
| 8 | Rio da areia | S8 | -14.635278 | 13.390556 |
| 9 | Montipa | S9 | -14.648889 | 13.260000 |
| 10 | Residence 1 | S10 | -14.759400 | 13.349360 |
| 11 | Residence 2 | S11 | -14.760150 | 13.351490 |
| 12 | Residence 3 | S12 | -14.746890 | 13.353420 |
| 13 | Residence 4 | S13 | -14.748180 | 13.313120 |
| 14 | MukwahonaResidence | S14 | -14.783600 | 13.352000 |
| 15 | Humbia | S15 | -14.760700 | 13.402900 |
| 16 | Navindombo | S16 | -14.733400 | 13.402500 |

In our study the water samples were taken twice, one in the Summer of 2017 and the other in the same period in 2018. The samples were analyzed with the RAD7 equipment and the radon concentration activities are shown in Table 3. In Figure we can see that 2 of the collected samples have values over 100 Bq L$^{-1}$, the reference value recommended in the directive 2013/EURATOM of the 22nd of October 2013 [19].

Table 3 presents the mean values of radon concentration from samples collected in the two years specified (S1 to S10). The highest obtained value from water samples was 207 ± 14 Bq L$^{-1}$ while the lowest obtained value was 39 ± 6 Bq L$^{-1}$. The values of the other water samples (S11 to S16) are presented in Table 4. From this table we can see that the mean value of radon concentration activity is 82.85 ± 8.8 Bq L$^{-1}$ in the sample collected during the summer of 2017 and 86.5 ± 8.7 Bq L$^{-1}$ for the samples collected in the summer of 2018. These values were used to build the graphic shown in Figure 4.

Table 3. Radon concentration activities of drinking water samples.

| Sample ID | Rn (Bq L$^{-1}$) Summer 2017 | Rn (Bq L$^{-1}$) Summer 2018 | Average (Bq L$^{-1}$) |
|---|---|---|---|
| S1 | 207 ± 14 | 197 ± 12 | 202 ± 13 |
| S2 | 100 ± 10 | 85 ± 9 | 92.5 ± 9.5 |
| S3 | 54 ± 7 | 61 ± 8 | 57.5 ± 7.5 |
| S4 | 119 ± 11 | 117 ± 11 | 118 ± 11 |
| S5 | 59 ± 7 | 48 ± 6 | 53.5 ± 6.5 |
| S6 | 66 ± 8 | 74 ± 8 | 70 ± 8 |
| S7 | 66 ± 8 | 73 ± 8 | 69.5 ± 8 |
| S8 | 75 ± 9 | 96 ± 10 | 85.5 ± 9.5 |
| S9 | 39 ± 6 | 40 ± 6 | 39.5 ± 6 |
| S10 | 83 ± 9 | 85 ± 9 | 84 ± 9 |

Table 4. Radon concentration activities of drinking water samples.

| Sample ID | Rn (Bq L$^{-1}$) Summer 2017 | Sample ID | Rn (Bq L$^{-1}$) Summer 2018 |
|---|---|---|---|
| S11 | 61 ± 8 | S14 | 79 ± 8 |
| S12 | 53 ± 7 | S15 | 87 ± 9 |
| S13 | 95 ± 10 | S16 | 83 ± 9 |
| Mean | 82.85 ± 8.8 | | 86.5 ± 8.7 |
| SD | 45.1 | | |

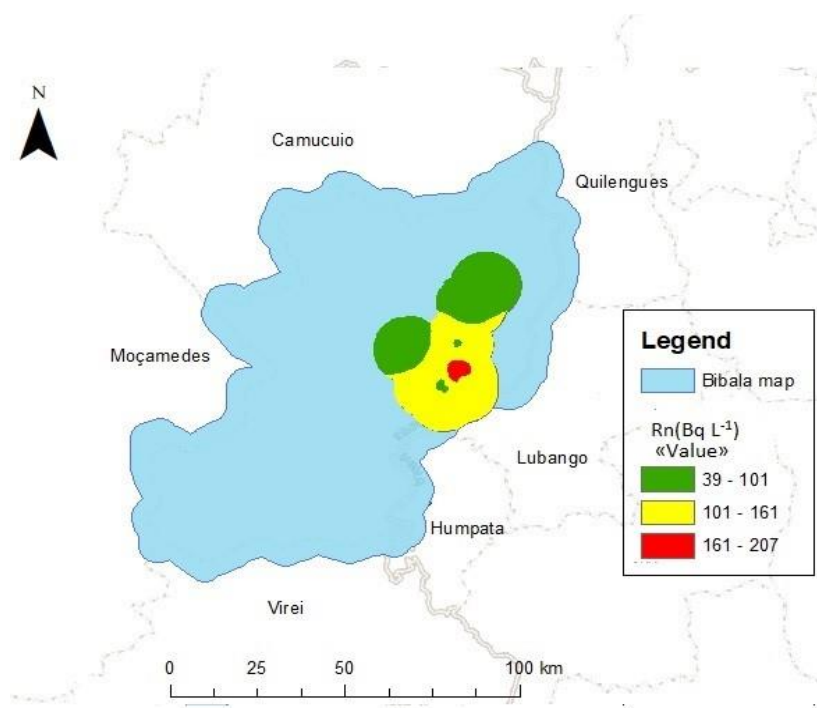

Figure 3. Radon concentration in water distribution at Bibala municipality.

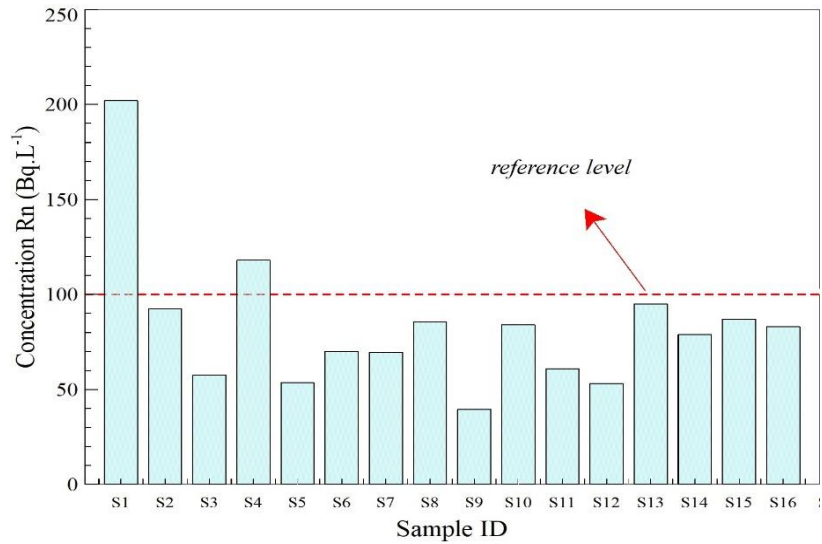

Figure 4. Distribution of radon concentration of the collected water samples.

The annual committed effective dose due to ingestion for individual water consumption of the collected samples in Bibala municipality, for the International Commission on Radiological Protection (ICRP) age groups, was evaluated using equation (1) and the results are shown in Table 5 and Figure 5. The annual committed effective dose due inhalation was evaluated using equation (3), and the results are shown in Table 6 and Figure 6. In Table 7 and Figure 7 we present the annual effective dose to internal organs, stomach and lungs, due to ingested water with radon, which is calculated with equations (1) and (3).

Table 5. Annual committed effective dose by ingestion for ICPR age groups.

| Sample ID | Rn (Bq L$^{-1}$) | GI (mSv y$^{-1}$) | GII (mSv y$^{-1}$) | GIII (mSv y$^{-1}$) |
|---|---|---|---|---|
| S1 | 202 | 2.121 | 1.414 | 1.010 |
| S2 | 92.5 | 0.971 | 0.648 | 0.463 |
| S3 | 57.5 | 0.604 | 0.403 | 0.288 |
| S4 | 118 | 1.239 | 0.826 | 0.590 |
| S5 | 53.5 | 0.562 | 0.375 | 0.268 |
| S6 | 70 | 0.735 | 0.490 | 0.350 |
| S7 | 69.5 | 0.730 | 0.487 | 0.348 |
| S8 | 85.5 | 0.898 | 0.599 | 0.428 |
| S9 | 39.5 | 0.415 | 0.277 | 0.198 |
| S10 | 84 | 0.882 | 0.588 | 0.420 |
| S11 | 61 | 0.641 | 0.427 | 0.305 |
| S12 | 53 | 0.557 | 0.371 | 0.265 |
| S13 | 95 | 0.998 | 0.665 | 0.475 |
| S14 | 79 | 0.830 | 0.553 | 0.395 |
| S15 | 87 | 0.914 | 0.609 | 0.435 |
| S16 | 83 | 0.872 | 0.581 | 0.415 |
| Mean | 83.1 | 0.873 | 0.582 | 0.416 |

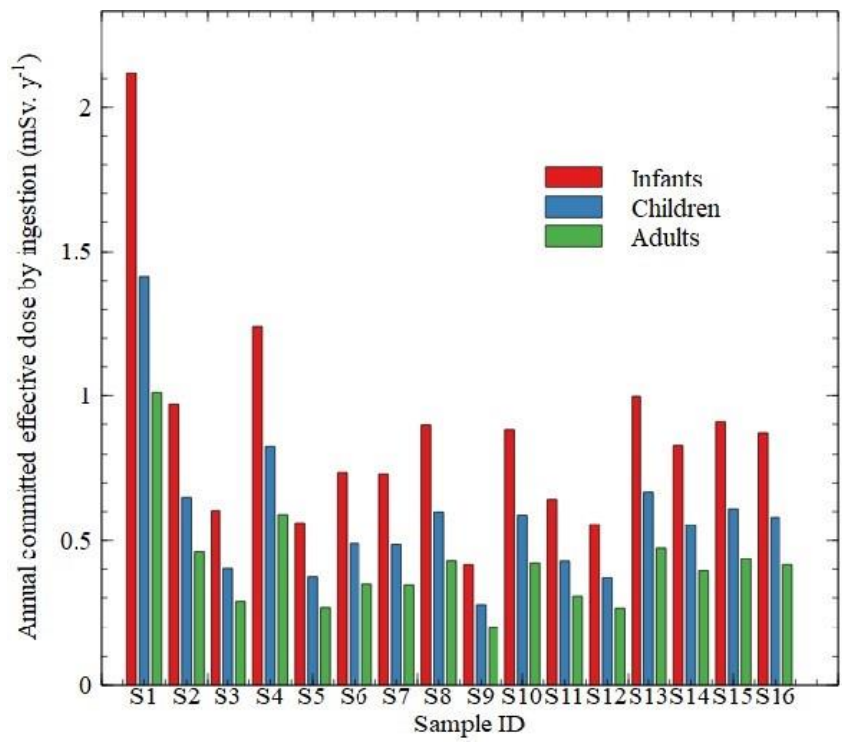

Figure 5. Annual committed effective dose due to ingestion.

Table 6. Annual committed effective dose by inhalation

| Sample ID | Rn (Bq L$^{-1}$) | Inhalation dose (mSv y$^{-1}$) |
| --- | --- | --- |
| S1 | 202 | 0.509 |
| S2 | 92.5 | 0.233 |
| S3 | 57.5 | 0.145 |
| S4 | 118 | 0.297 |
| S5 | 53.5 | 0.135 |
| S6 | 70 | 0.176 |
| S7 | 69.5 | 0.175 |
| S8 | 85.5 | 0.215 |
| S9 | 39.5 | 0.100 |
| S10 | 84 | 0.212 |
| S11 | 61 | 0.154 |
| S12 | 53 | 0.134 |
| S13 | 95 | 0.239 |
| S14 | 79 | 0.199 |
| S15 | 87 | 0.219 |
| S16 | 83 | 0.209 |
| Mean | | 0.209 |

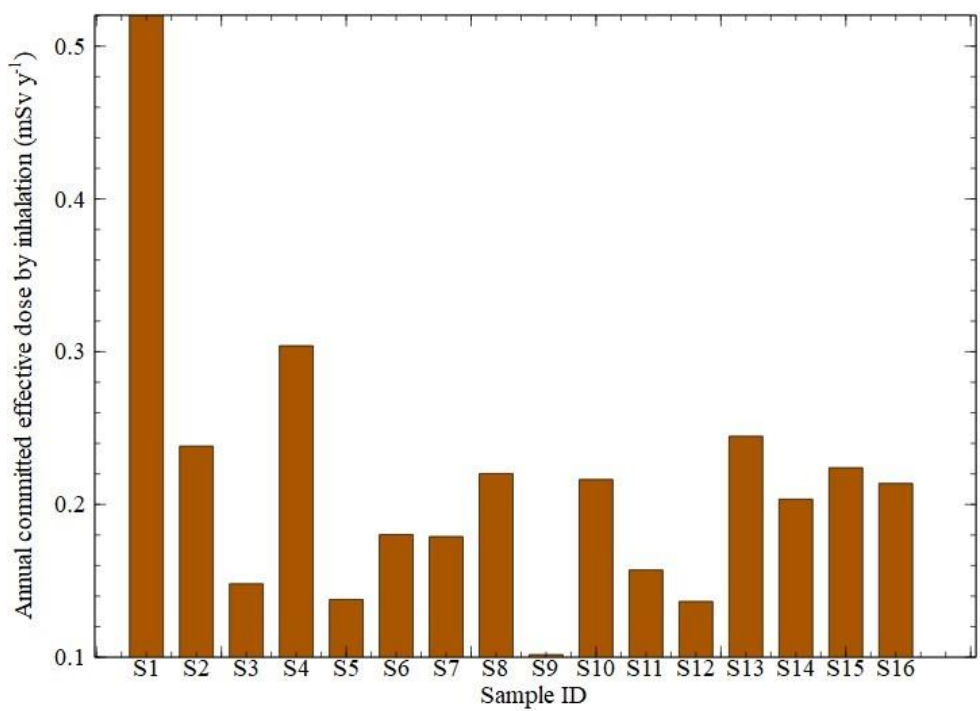

Figure 6. Annual committed effective dose due to inhalation.

Table 7. Annual committed effective dose to internal organs: stomach and lungs for different age groups.

| Sample ID | $D_{GI}$ (*mSv y*$^{-1}$) | $D_{GII}$ (*mSv y*$^{-1}$) | $D_{GIII}$ (*mSv y*$^{-1}$) |
| --- | --- | --- | --- |
| S1 | 2.630 | 1.923 | 1.519 |
| S2 | 1.204 | 0.880 | 0.696 |
| S3 | 0.749 | 0.547 | 0.432 |
| S4 | 1.536 | 1.123 | 0.887 |
| S5 | 0.696 | 0.509 | 0.402 |
| S6 | 0.911 | 0.666 | 0.526 |
| S7 | 0.905 | 0.661 | 0.523 |
| S8 | 1.113 | 0.814 | 0.643 |
| S9 | 0.514 | 0.376 | 0.297 |
| S10 | 1.093 | 0.799 | 0.632 |
| S11 | 0.794 | 0.581 | 0.459 |
| S12 | 0.690 | 0.505 | 0.398 |
| S13 | 1.237 | 0.904 | 0.714 |
| S14 | 1.029 | 0.752 | 0.594 |
| S15 | 1.133 | 0.828 | 0.654 |
| S16 | 1.081 | 0.790 | 0.624 |
| Mean | 1.081 | 0.791 | 0.625 |

Drinking water with radon is a natural source of radiation dose to the human body through ingestion and through inhalation. In this study, the mean value of the estimated dose to the public in the selected area by ingestion is 0.873 mSv y$^{-1}$, 0.582 mSv y$^{-1}$ and 0.416 mSv y$^{-1}$ for infants, children and adults, respectively and an average value of 0.209 mSv y⁻¹ for dose to inhalation. From the analysis of Table 7, we can conclude that radiation average dose to the stomach and lungs is 1.081 mSv y⁻¹, 0.791 mSv y⁻¹ and 0.625 mSv y⁻¹ for infants, children and adults, respectively. Analysing these results, we can clearly see that the dose received by the stomach due to ingestion is lower than the dose received by the lungs. When radon is transferred from the water to the air, possibly because the different radio sensibility of these organs.

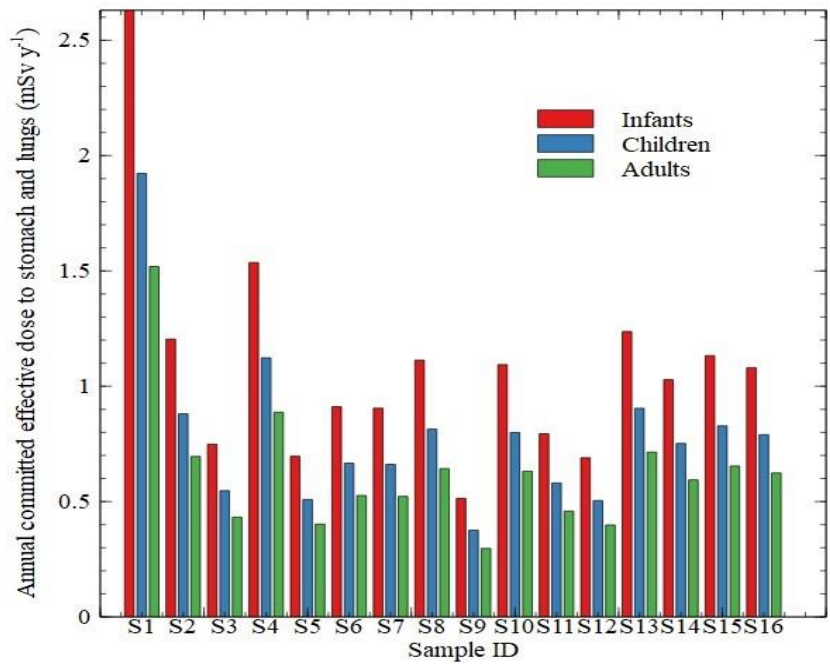

Figure 7. Total effective dose due to ingestion and Inhalation.

4. Conclusions

Granitic rocks are quite common in the Southern region of Angola and it is also frequently found that radioactive minerals containing, for instance, $^{238}$U, that has radon on its decay chain. Radon short-lived progeny $^{218}$Po and $^{214}$Po that are solid, when inhaled or ingested, constitute a source of exposure to radiation that may cause undesirable health effects.

This was the main motivation to measure the radon concentration in the drinking water for Bibala municipality. Looking at the results obtained from measurements performed on 16 samples, we ascertain that 2 of those samples are above the safety limits set by WHO (100 Bq L⁻¹) and all are above the limits set by US-EPA (11 Bq L⁻¹). However, considering the action level recommended by the European Directive 2013/51/EURATOM (100 Bq L⁻¹), we see that 2 of the samples have values over this limit. Additionally, it was observed that, for all the samples in the study area, the total effective dose from radon-in-water is higher than the safety the limits set for the public of 0.1 mSv y⁻¹ suggested by ICRP [20].

As far as we know, this work of measuring radon concentration in drinking water, is the first study done in Bibala. Even though the water may not be safe for immediate consumption, the reported values don't show the need for an immediate preventive action. Thus, the obtained results can be a baseline data for radon measurements in water. However, it is still necessary to carry out more work on radon activity measurements in Bibala municipality, increasing the number of analyzed samples and, if it possible, extend the work across the country.

Acknowledgements

We acknowledge the support of this work by the government of the Republic of Angola through PhD grants n° 38/2016 and n° 441/2016 for J. Kessongo and Y. Bahu, respectively. We are grateful to Ashley Peralta for the review of the English text.